\documentclass[12pt]{iopart}

\usepackage{amsfonts}
\usepackage{amssymb}
\usepackage{color}
\usepackage{units}
\usepackage{verbatim}
\usepackage{graphicx}
\usepackage[normalem]{ulem} 

\newcommand{\beq}{\begin{equation}}
\newcommand{\eeq}{\end{equation}}
\newcommand{\bqa}{\begin{eqnarray}}
\newcommand{\eqa}{\end{eqnarray}}

\newcommand{\erf}[1]{Eq.~(\ref{#1})}

\newcommand{\bra}[1]{\left\langle{#1}\right|}
\newcommand{\ket}[1]{\left|{#1}\right\rangle}

\newcommand{\sch}{Schr\"odinger}

\newcommand{\sq}[1]{\left[ {#1} \right]}

\newcommand{\an}[1]{\left\langle{#1}\right\rangle}

\newcommand{\st}[1]{\left| {#1} \right|}
\newcommand{\s}[1]{\hat\sigma_{#1}}
\newcommand{\tp}{^{\top}}

\definecolor{ngreen}{rgb}{0.1,0.5,0.1}

\definecolor{golden}{rgb}{0.8,0.6,0.1}

\definecolor{purp}{rgb}{0.8,0.1,0.8}

\definecolor{orange}{rgb}{0.9,0.3,0}
\definecolor{mar}{rgb}{0.6,0.1,0.1}

\renewcommand{\Tr}[1]{{\rm Tr}\sq{ {#1} }}

\begin{document}

\title[The Simplest Demonstrations of Quantum Nonlocality]{The Simplest Demonstrations of Quantum Nonlocality}

\author{Dylan J. Saunders$^1$, Matthew S. Palsson$^1$, Geoff J. Pryde$^{1,*}$, Andrew J. Scott$^1$, Stephen M. Barnett$^{2,\dagger}$ and Howard M. Wiseman$^{1,\ddagger}$}

\address{$^1$ Centre for Quantum Computation and Communication Technology (Australian Research Council),  Centre for Quantum Dynamics, Griffith University, Brisbane, 4111, Australia}
\address{$^2$ Department of Physics, University of Strathclyde, Glasgow G4 0NG, U.K.}

\ead{$^*$G.Pryde@griffith.edu.au}
\ead{$^\dagger$Steve@phys.strath.ac.uk}
\ead{$^\ddagger$H.Wiseman@griffith.edu.au}

\begin{abstract}
We investigate the complexity cost of demonstrating the key types of nonclassical correlations --- Bell inequality violation, EPR-steering, and entanglement --- with independent agents, theoretically and in a photonic experiment. We show that the complexity cost exhibits a hierarchy among these three tasks, mirroring the recently-discovered hierarchy for how robust they are to noise. For Bell inequality violations, the simplest test is the well-known CHSH test, but for EPR-steering and entanglement the tests that involve the fewest number of detection patterns require non-projective measurements. The simplest EPR-steering requires a choice of projective measurement for one agent and a single non-projective measurement for the other, while the simplest entanglement test uses just a single non-projective measurement for each agent. In both of these cases, we derive our inequalities using the concept of circular 2-designs. This leads to the interesting feature 
that in our photonic demonstrations, the correlation of interest is independent of the angle between the linear polarizers used by the two parties, which thus require no alignment.

\end{abstract}


\maketitle

\section{Introduction}

Nonclassical correlations are a powerful resource for information processing, and studying them opens a window on the true nature of the quantum world. For example, entanglement appears to be a requirement for universal quantum computing, while violations of Bell inequalities reveal that it is not possible to describe certain physical phenomena using  any local realistic model \cite{BelPHY64}.

Recently there has been significant interest in the hierarchy of nonclassical correlations. It has been known for some time that not all entangled states are capable of violating a Bell inequality \cite{Werner1989}. In 2007, the EPR-steering phenomenon was added to this ladder of effects. EPR-steering is a formalization \cite{Wiseman2007,Jones2007} of Einstein's ``spooky action at a distance''--- the Einstein, Podolsky, Rosen (EPR) effect \cite{EinEtalPR35,RevModPhys-EPRparadox} ---  called ``steering" by \sch\ \cite{SchPCP35}. This formalization was achieved by turning EPR-steering into a real-world quantum information {\em task}, in which  mixed states and arbitrary (non-projective) measurements must be considered~\cite{Wiseman2007,Jones2007}. \sch's term is appropriate because this task can be performed only if one party (Alice) demonstrates that she can steer the state of the other party's (Bob's) sub-system, via her choice of her measurement setting. That is, EPR-steering is demonstrated if and only if Bob is forced to admit that he cannot describe his system by a local quantum state unaffected by Alice's actions. Note that while Bob's measurement apparatus is assumed trustworthy, just as in entanglement tests, no assumptions are made about the apparatus that generates Alice's declared outcomes, just as in Bell tests.

There is a hierarchy of states that can be used to demonstrate these {non-local} phenomena: the set of states that can be used to demonstrate Bell-nonlocality is a strict subset of those that can manifest EPR-steering, which is likewise a strict subset of those showing entanglement \cite{Wiseman2007,Jones2007}. Very recently this hierarchy has been demonstrated experimentally using a range of two-qubit Werner states, and utilising projective measurements \cite{Saunders2010}. It has also been extended theoretically to higher dimensional spin systems \cite{Reid2011}. The hierarchy of nonclassical correlations also extends to the quantum information protocols they enable. For example, a Bell inequality violation can provide device-independent security for quantum key distribution --- Alice and Bob can be certain to share a secure key, even if each of their apparatus was provided by an eavesdropper \cite{Acin2007}. EPR-steering inequalities, on the other hand, enable an asymmetric form of this security in which only 
Bob's detector need be trusted \cite{Branciard2011}. In both cases, closing the detection loophole is essential, and this was recently demonstrated for EPR-steering in a number of photonic experiments \cite{Bennet2011,Smith2012,Wittmann2012}\footnote{
Interestingly, the first of these \cite{Bennet2011} showed that, in the absence of mixture, there is no strict hierarchy for loss tolerance, in the sense that
EPR-steering {\em with no detection loophole} can be demonstrated with arbitrarily low detector efficiency for Alice as well as Bob, 
just as for entanglement witnessing.}.

In this paper, we uncover the hierarchy of nonclassicality from a different perspective. Rather than asking what degree of purity of the state is required, or what tasks are enabled, we ask how complex an experiment is required for a given demonstration. As well as establishing a hierarchy of complexity cost, this has practical application in maximizing the bandwidth in quantum information protocols, such as entanglement distribution, that require non-local correlations.

To address this problem, we define the fundamental complexity ``cost'' by the integer $W$, the number of different types of detection patterns (joint detection outcomes) that must occur. We find that this measure reproduces the same strict hierachy: The most parsimonious (least complex) demonstrations of entanglement, EPR-steering, and Bell nonlocality require $W=9$, $12$, and $16$ respectively. While the most parsimonious test of Bell nonlocality is the standard CHSH \cite{CHSH:1969} test using projective measurements, those for entanglement and EPR-steering both require generalized (non-projective) measurements \cite{Barnett2009}. We implement these maximally parsimonious demonstrations experimentally using polarization-entangled photons.  The demonstrations of entanglement and EPR-steering have the interesting feature that they do not require alignment of the polarizers between the parties: the degree of violation of the corresponding inequalities is independent of the relative angle. 

This paper is structured as follows. In Sec.~2 we define the problem precisely and show from very general principles that the costs for demonstrating entanglement, EPR-steering, and Bell nonlocality cannot be lower than $W=9$, $12$, and $16$ 
respectively. In Sec.~3, we introduce a formalism, using the concept of 1-designs and 2-designs, to describe certain sorts of measurements on an entangled pair of qubits. This allows us, in Sec.~4, to construct new nonlocality tests that attain the stated minimum $W$ values for entanglement tests and EPR-steering. Finally (Sec.~5), using photonic singlet states, we demonstrate experimentally the violation of these inequalities using ``trine'' measurements \cite{Barnett2009,ClarkePRA:2001}, as well as of the standard CHSH inequality,

\section{Defining the problem} 

Consider an experimental demonstration of some particular type of quantum nonlocality involving $P\geq 2$ distinct parties. Let $S_p\geq 1$ be the number of different measurement settings used by party $p$, and $O_{s_p}\geq 2$ the number of possible outcomes of setting $s_p$ for party $p$. Then the definition of the complexity cost is 
 \beq \label{defW}
W = \prod_{p=1}^{P} \sum_{s_p=1}^{S_p} {O_{s_p}}.
\eeq 
In words, $W$ is the number of possible patterns of joint detection outcomes that can occur. In the remainder of this paper, we restrict to $P=2$, and name the two parties Alice and Bob.  For $P>2$ the minimum complexity cost  is certainly never smaller than for $P=2$, but we note that it is an interesting open question to consider the minimum $W$ required to demonstrate   ``genuine multipartite quantum nonlocality'' (see e.g. \cite{Bancal2011}) of various kinds.

Note that we have taken the choice of setting by each party to be independent. Dropping this assumption allows more parsimonious tests than the ones we consider, but at the cost of lowering security. Both entanglement and EPR-steering could be demonstrated by Alice and Bob with qubits and projective measurements, with only eight possible patterns: they can measure the correlation $\an{\s{x} \otimes \s{x}+\s{z}\otimes \s{z}}$ if either both measure $\s{x}$, or both measure $\s{z}$. 
In the absence of communication between the parties, this would require Alice and Bob to have predetermined which of $\s{x}$ or $\s{z}$ they measure in each run. As stated, however, such predetermination opens a loophole in the tests, as we now explain for the two cases. For the case of EPR-steering, a rigorous test means that Alice is not trusted by Bob \cite{Wiseman2007}. Clearly if Bob's settings are predetermined then Alice could trivially create a perfect correlation by sending Bob either $\s{x}$ eigenstates or $\s{z}$ eigenstates as appropriate. For the simpler case of demonstrating  entanglement both parties are considered trustworthy \cite{Wiseman2007}. However, if the device supposedly creating the entanglement is untrusted --- it might be supplied by a competitor, Eve --- then the device could spy on the predetermined lists of settings and again create perfect correlations while generating only product states. For this reason we stick to the rigorous conditions in which the parties' choices are independent, giving \erf{defW}.

We first show that the values of $W=16$, $12$, and $9$ are the minimum possible to demonstrate Bell nonlocality, EPR-steering, and  entanglement respectively. To demonstrate Bell nonlocality it is necessary for both Alice and Bob to have measurement choices; without this it is trivial to construct a local-hidden-variable model that explains the measurement outcomes. Thus the most parsimonious inequality is that found originally by Bell, and refined by CHSH \cite{CHSH:1969}, in which each party has two settings, and each setting has two outcomes. The Bell-nonlocality case therefore has $W_{\rm B}=16$. Next, for EPR-steering it is again crucial for Alice to have a measurement choice (the task is for her to steer Bob's state by her choice of measurement), but Bob need not use more than one setting. However, if he uses only one setting then he cannot use a two-outcome measurement because the two positive operators corresponding to the two outcomes would necessarily commute, and hence be co-diagonal in some basis. 
Completely decohering Bob's system in that same basis would have no effect on his outcomes, but would of course completely destroy any entanglement with Alice's system and hence any possibility of demonstrating EPR-steering. Thus the smallest cost $W$ in this case is when Alice has two settings with two outcomes each, and Bob has one setting with three outcomes, giving $W_{\rm S} = 12$. Finally, for demonstrating entanglement no choice is required on either side, but with this strategy it is necessary to have three-outcome measurements on both sides, using the  same commutation argument, yielding a minimum of $W_{\rm E} = 9$.   We also note that the heirarchy is preserved with respect to the number of measurement \textit{settings}, i.e.\ the number of distinct configurations of the measurement apparatus. Demonstrations of the CHSH, EPR-steering and entanglement phenomena require 4, 2, and 1 measurement settings respectively.

\section{POVMs, spherical designs, and the singlet correlations}

Consider an arbitrary  $N$-outcome qubit POVM (positive operator valued measure; also called a POM \cite{Barnett2009}), $\{\hat F_k\}_{k=1}^N$ which is {\em sharp} \cite{MarMuy90,WisMil10}.  In other words, each $\hat F_k$ is proportional to a rank-one projector on the qubit Hilbert space.  Each POVM element $\hat F_k$ can thus be defined by a unit-length 3-vector $\vec{A}_k$, and a positive weight $a_k$ by 
\beq
\hat F_k = a_k (\hat{1}+\vec{A}_k\cdot \vec\sigma).
\eeq
Here $\hat{1}$ is the qubit identity operator and $\vec\sigma = (\s{x},\s{y},\s{z})\tp$ is the vector of Pauli operators. The POVMs have the completeness condition that $\sum_{k=1}^N \hat F_k = \hat 1$, which implies  
 that  
\beq
\sum_{k=1}^N a_k = 1,\;\; \sum_{k=1}^N  a_k \vec{A}_k = \vec{0}. \label{spherical1design}
 \eeq
This means that the set $\mathbb{A} = \{a_k, \vec{A}_k \}_{k=1}^N$ forms a $1$-design \cite{Delsarte:1977}.  Taking the above to apply to Alice's qubit, we likewise define POVM elements for Bob by $\hat E_j = b_j (\hat{1}+\vec{B}_j\cdot \vec\sigma)$, with the analogous constraints on his set $\mathbb{B} = \{b_j, \vec{B}_j \}_{j=1}^M$. We will use $\mathbb{A}$ and $\mathbb{B}$ to denote Alice's and Bob's measurement.

Say Alice and Bob share the 2-qubit singlet state $$\rho_{\rm  singlet} =  (1/4)({\hat{1}\otimes\hat{1}}- \s{x}\otimes \s{x}-\s{y}\otimes \s{y}-\s{z}\otimes \s{z}).$$ From the Pauli algebra it is easy to verify that the probability for the joint outcome $(k,j)$ is 
\beq \label{Pkj}
P_{kj} = \Tr{\rho_{\rm singlet} (\hat F_k \otimes \hat E_j)} = a_kb_j (1-\vec{A}_k\cdot \vec{B}_j).
\eeq Note that we are using Tr to denote a trace over a quantum Hilbert space; we reserve tr to denote a trace over 
a tensor relating to 3-dimensional space (see below). 

We wish to consider correlations between Alice's and Bob's outcomes. So far we have only labels $k$ 
and $j$ for these outcomes, but no physical reason for looking at any particular relation between them. We could look at the entire (discrete) probability distribution $P_{kj}$, as is done in exhaustive searches for Bell-inequality violations \cite{BrunGis08}. This is appropriate for investigating Bell-nonlocality, a concept which makes no reference to quantum mechanical properties of the systems. Here, however, we are interested in demonstrating other sorts of nonlocality in which such properties do play a role, namely entanglement and EPR-steering. Thus we seek a natural correlation function which uses the quantum physics of the problem. Rather than considering the outcome of Alice's measurement ${\mathbb A}$ to be the label $k$, we will consider it to be a unit vector $\vec{A}$, taking values $\vec{A}_k$ corresponding to the POVM elements $\hat F_k$, and likewise for Bob. 
Then considering \erf{Pkj}, the natural way to combine the two vector-valued random variables $\vec{A}$ and $\vec{B}$ is to construct the correlation function 
\beq \label{natcf}
\an{\vec{A}\cdot \vec{B}} =
\sum_{jk} P_{kj} \vec{A}_k\cdot \vec{B}_j.
\eeq

Using the fact that ${\mathbb A}$ and ${\mathbb B}$ are spherical 1-designs, 
and the singlet correlations (eq.~\ref{Pkj}),  
we find that the ensemble average (\ref{natcf}) evaluates to the simple expression
\bqa
\an{\vec{A}\cdot \vec{B}} 
&=& -\tr\sq{{\bf A}{\bf B}}, \label{tensorform}
\eqa
where we have defined the $3\times3$  tensor 
\beq 
{\bf A} = \sum_{k=1}^N a_k \vec{A}_k \vec{A}_k\tp,
\eeq 
and similarly for ${\bf B}$. Note that $\tr{\bf A} = \tr{\bf B} = 1$ from normalization.

Now if, in addition to \erf{spherical1design}, ${\mathbb A}$ has the property that
 ${\bf A} = (1/3){\bf I}_3$, where ${\bf I}_3$ is the $3\times 3$ identity tensor, then  
${\mathbb A}$ forms a weighted spherical 2-design on $S^2$ \cite{Hong:1982}. 
The concept of spherical $t$-designs is defined for all $t\in {\mathbb N}$; 
intuitively, as $t$ increases this means that ${\mathbb A}$ is a better and better approximation to the set of all unit vectors on $S^2$, distributed according to the $SO(3)$-Haar measure. In this paper 
we are interested only in 1-designs and 2-designs. The simplest  example of a spherical 2-design on the 2-sphere involves four equally-weighted vectors $\vec{A}_j$ pointing to the vertices of a regular terahedron. 

Say that ${\bf B} = (1/3) {\bf I}_3$, so that ${\mathbb B}$ is a weighted 2-design on the sphere. Then
 for any 1-design (that is, any measurement) ${\mathbb A}$, it follows from \erf{tensorform} that 
\beq \label{singlet-sphere}
\an{\vec{A}\cdot \vec{B}} = -\tr{\bf A} /3 = -1/3.
\eeq
That is, the correlation between Alice's and Bob's measurement results is {\em independent}
of the weightings and 
3-dimensional orientation of the  vectors that define their POVMs. This supports the idea that 
$\an{\vec{A}\cdot \vec{B}}$ is a natural correlation function for qubit POVMs.
 
\begin{figure}\begin{center}
\includegraphics[width=.5\linewidth]{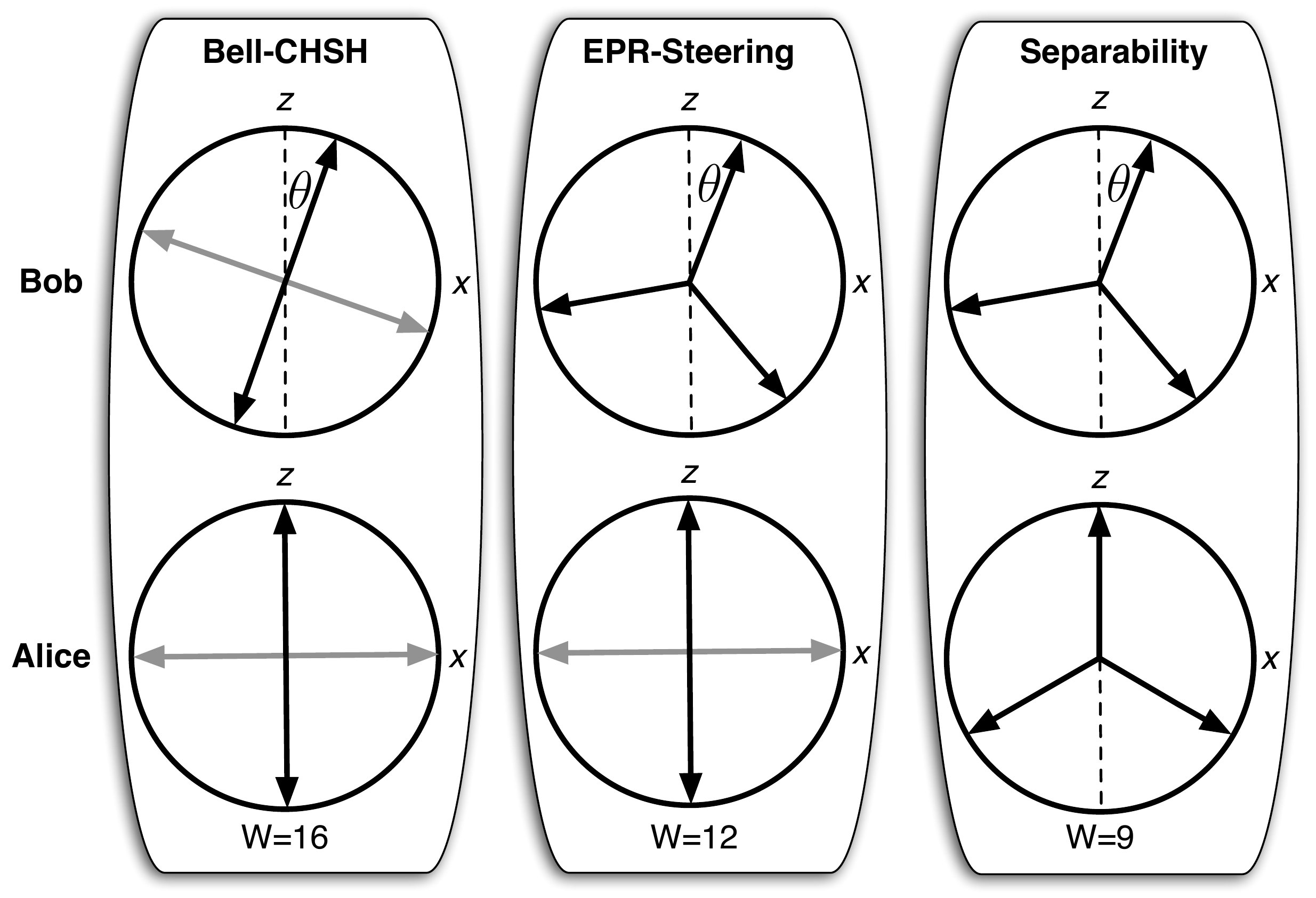}\end{center}
\vspace{-3ex} \caption{Measurement Directions. The circle represents the $x-z$ plane of the Bloch sphere, the dashed line shows the $z$ axis, vectors represent the directions $\vec{A}$ and $\vec{B}$, and different shades indicate different measurement settings. $\theta$ is the angle between the primary measurement axes of Alice and Bob. We vary $\theta$ to test the rotational (in)variance of the three correlation functions.}
\label{fig:Measurements}\end{figure}

To derive the most parsimonious tests of EPR-steering and entanglement we actually wish to consider 
 not spherical 2-designs, but circular $2$-designs \cite{Hong:1982}. This applies if we restrict the $\vec{A}_k$ and $\vec{B}_j$ to lying in 
 a single plane in the Bloch sphere (say the $y=0$ plane). Then ${\mathbb B}$ being a  circular $2$-design means $\sum_{j=1}^M b_j \vec{B}_j \vec{B}_j\tp = (1/2){\bf I}_2$ where ${\bf I}_2$ is the $2\times 2$ identity tensor (in the $x$--$z$ plane). The simplest example of a circular 2-design involves three equally weighted vectors $\vec{B}_j$ pointing to the vertices of an equilateral triangle in the $x$--$z$ plane (see Fig.~\ref{fig:Measurements}). The corresponding measurement is known as a trine measurement \cite{ClarkePRA:2001,Barnett2009}. For any circular 2-design measurement by Bob, in which the $\vec{A}_k$ are restricted to the same plane, \erf{singlet-sphere} is replaced by 
\beq 
\label{singlet-plane}
\an{\vec{A}\cdot \vec{B}} = -\tr{\bf A} /2 = -1/2.
\eeq 
It is still the case that the singlet correlation between Alice's and Bob's measurement results is independent of the weightings and planar  arrangement of the  vectors, only now the degree of correlation is higher.

From the discussions in Sec.~2 we know that the simplest possible tests of entanglement or EPR-steering would involve two- or three-outcome measurements. Henceforth we now restrict, for simplicity, to the case where Bob makes a trine measurement, and Alice either a trine measurement or two orthogonal projective measurements. Both sides' measurements are assumed to lie in the same plane (e.g. linear polarization measurements) but we make no assumptions about the relative angle between Alice's and Bob's designs.

\section{Quantum nonlocality tests} 
We first derive an entanglement test. For any single qubit state $\ket{\psi}$  with $\mathbb{B}=$ trine, 
\beq
\st{\an{\vec{B}}}  = \st{\sum_{j=1}^{M} \vec{B}_j\bra{\psi} \hat E_j\ket{\psi} }  \leq  \frac{1}{2} , \label{Bob-LHS}
\eeq
The proof is similar to that for the singlet correlations, \erf{singlet-plane}. Thus, if Alice and Bob have a separable state, and both make trine measurements, then 
\beq \label{ent_wit}
\st{\an{\vec{A}\cdot \vec{B}}} \leq \frac{1}{2} \times \frac{1}{2} = \frac{1}{4}.
\eeq
But for the singlet state, using \erf{singlet-plane}
\beq \label{pars-ent}
\st{\an{\vec{A}\cdot \vec{B}}} = \frac{1}{2} > \frac{1}{4}.
\eeq
Thus,  \erf{ent_wit} can be violated, and so is a useful witness of entanglement, attaining the minimum $W_{\rm E}=9$.

We can derive an EPR-Steering inequality \cite{CavJonReiWis09} in a very similar way, but with Alice using two orthogonal (in the Bloch sphere sense) projective measurements, ${\mathbb A}$ and ${\mathbb A}'$. We thus have, for any pair of Alice's results, 
\beq
\st{\vec{A} + \vec{A}' } \leq 2\cos\frac{\pi}{4} = \sqrt{2}.
\eeq
Note that this makes no use of quantum mechanics, but is a property only of the vectorial values of the results Alice reports.  This is as required for a rigorous demonstration of EPR-steering \cite{Wiseman2007}. But Bob still trusts his apparatus so under the assumption that Bob has a local quantum state unaffected by Alice's choice of setting,  \erf{Bob-LHS} still applies \cite{Wiseman2007}. Thus for $\mathbb{B}=$ trine we can derive \beq
\st{\an{(\vec{A} + \vec{A}') \cdot\vec{B} }}  \leq \sqrt{2} \times \frac{1}{2} = \frac{1}{\sqrt{2}}. 
\label{steering-ineq}
\eeq
For a singlet state, however, we have from \erf{singlet-plane}, 
\beq
\st{\an{(\vec{A} + \vec{A}')\cdot \vec{B} }} = 1 > \frac{1}{\sqrt{2}}
\eeq
Thus, \erf{steering-ineq} is an EPR-steering inequality \cite{CavJonReiWis09} that can 
be violated and attains the minimum value of  $W_{\rm S} = 12$.

We note finally that the CHSH inequality  
\beq |\an{AB + A'B + AB' - A'B'}| \leq 2, \label{CHSH}
\eeq
which achieves the minimum cost of $W_{\rm B}=16$, is not of the form that we considered for entanglement and EPR-steering; the variables $A$, $A'$, $B$ and $B'$ are not unit-vectors but rather take the values $\pm 1$. 

\section{Experiment} 
\begin{figure}\begin{center}
\includegraphics[width=.7\linewidth]{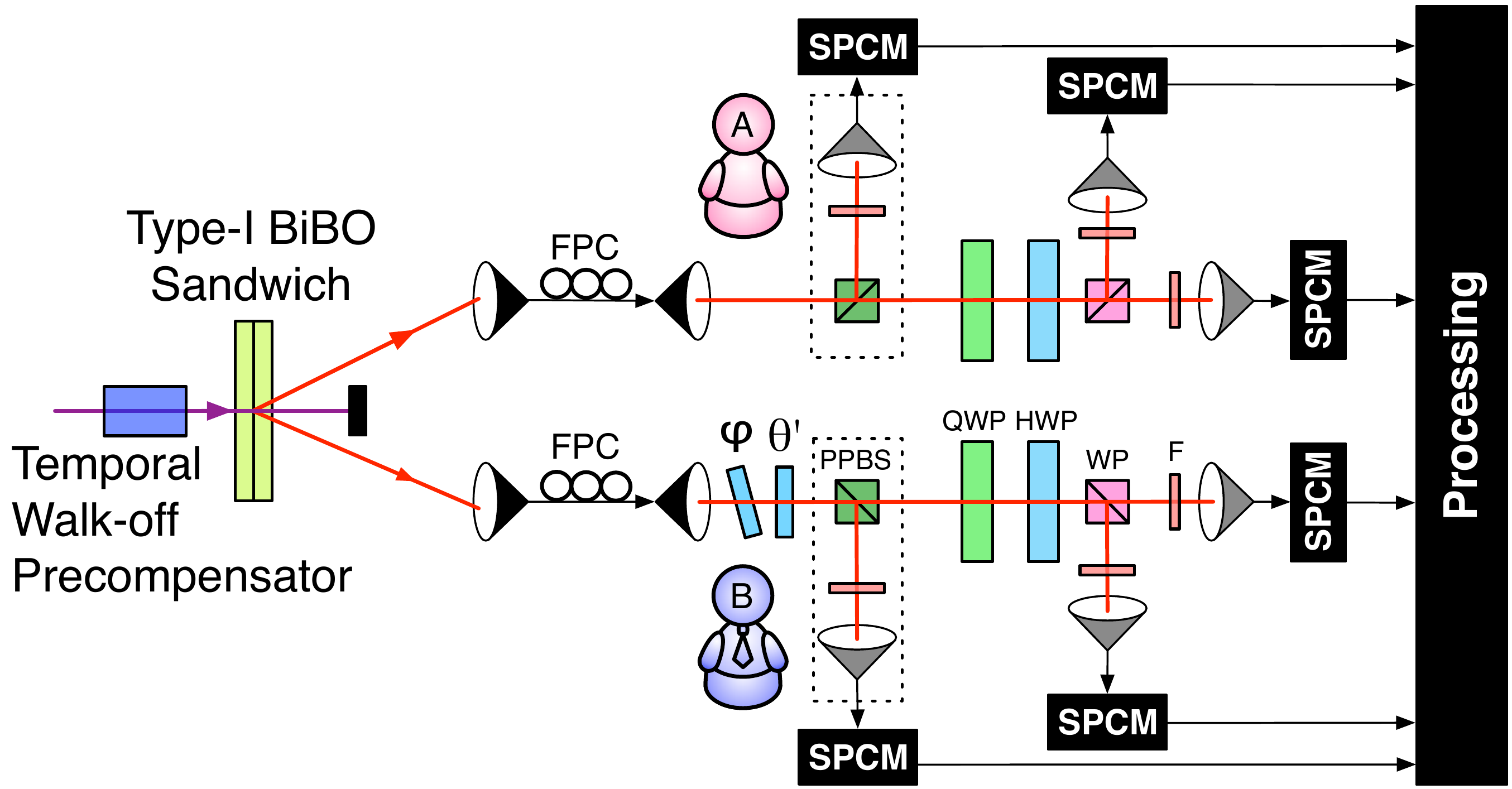}\end{center}
\vspace{-3ex} \caption{Experimental Setup (Color online). Alice's lab is the top rail, while Bob's is the bottom. A frequency-doubled mode-locked 820nm Ti-Sapphire laser with a 80MHz repetition rate drives a type-I parametric down converter in the sandwich configuration \cite{Kwiat:2005}, coupled into single mode fibres (black). Fibre polarisation controllers (FPC), combined with a phase shift $\varphi$, rotates the entangled state to the desired singlet state. The CHSH measurement schemes shown in Fig.~\ref{fig:Measurements} are implemented using half wave plates (HWP), quarter wave plates (QWP) and polarising Wollaston prisms (WP). Partially polarising beam splitters (PPBS) (dotted boxes) are inserted into the beam path when needed to implement the trine measurements. The photons are filtered with 3 nm FWHM filters (F) and then collected via multi mode fibres (grey) onto single photon counting modules (SPCM).} 
\label{fig:ExpSetup}
\end{figure}

We experimentally demonstrated the maximally parsimonious tests defined above using  photonic qubits, employing the polarization encoding $\ket{H}=\ket{0}$ and $\ket{V}=\ket{1}$. The entangled photons are generated from a type-I `sandwich' bismuth borate (BiBO) spontaneous parametric down conversion source \cite{Kwiat:2005}.  The entangled state produced is the maximally entangled Bell state $\ket{\Phi^{+}}=(\ket{00}+\ket{11})/\sqrt{2}$. We rotate $\ket{\Phi^{+}}$ to the desired singlet state state, $\ket{\Psi^{-}} = (\ket{01}-\ket{10})/\sqrt{2}$, using standard polarising optics; see Fig.~\ref{fig:ExpSetup} for details. The apparatus can be readily reconfigured to demonstrate each non-locality test.   We  also perform quantum state tomography \cite{James:2001}, which allows us to  determine the quality of the experimentally produced states.   

The entangled state  we produce has a fidelity of $(97 \pm 1)\%$ with the ideal  singlet, allowing us to easily violate all of our inequalities. The slight infidelity is caused by three effects, each making an approximately equal contribution. Imperfect collection of correlated photons produces a small amount of symmetric (depolarization) noise. Imperfect temporal walk-off compensation causes dephasing noise, which is not symmetric. Imperfect compensation of unwanted birefringence leads to a small local unitary rotation of the state, away from the symmetric singlet state. After allowing for the local unitary rotation, our experimental state has a fidelity of 99\% with the closest Werner state \cite{Werner1989}.

Each quantum non-locality inequality requires different optical elements. To demonstrate Bell non-locality, by violating a CHSH inequality \cite{CHSH:1969}, only projective measurements are needed. Demonstrating maximally parsimonious EPR-Steering and Entanglement witnesses requires at least one non-projective measurement. We simultaneously monitored all measurement outcomes, whilst calibrating for different channel efficiencies; see Fig.~\ref{fig:ExpSetup} and Appendix A.

We implement the non-projective trine measurement using the techniques of Ref.~\cite{ClarkePRA:2001}. The three POVM elements $\hat E_j$ are proportional to 
projectors onto the following three states, for the case $\theta = 0$ (See Fig.~\ref{fig:Measurements}),
\bqa
\ket{T_{0}} = \ket{1}\nonumber ;~
\ket{T_{\pm}} = -\frac{\sqrt{3}}{2}\ket{1}\pm\frac{1}{2}\ket{0}\nonumber.
\eqa
Each of these states is separated by an angle of $\frac{2\pi}{3}$ in the $x$--$z$ plane of the Bloch sphere. The weighting $b_j$ or $a_j$ appearing in each POVM element is $\frac{1}{3}$. To implement such a measurement, a partially projecting polarising element is added to the set-up to allow us to produce the required projectors with the appropriate weights. Specifically, we employ a PPBS with transmissivities $\tau_V=\sqrt{1/3}$ and $\tau_H=1$, and reflectivities $r_V=\sqrt{2/3}$ and $r_H=0$ . This allows us to implement the measurement $\bra{T_0}$ at the reflecting port  of the PPBS. A measurement of $\sigma_x$ is then performed  using a HWP at $\frac{\pi}{8}$ and a Wollaston prism on the transmitted port; the two outputs represent the outcomes $\bra{T_+}$ and $\bra{T_-}$. For those cases where a two-outcome projective measurement is required (rather than a trine measurement), the PPBS is omitted. In all cases, Bob can rotate the orientation of his measurement setting by the angle $\theta$ shown in Fig.~\ref{fig:Measurements} by using the HWP labelled $\theta^{'}$ ($\theta^{'}/4=\theta$) in Fig.~\ref{fig:ExpSetup}.

\begin{figure}\begin{center}
\includegraphics[width=.6\linewidth]{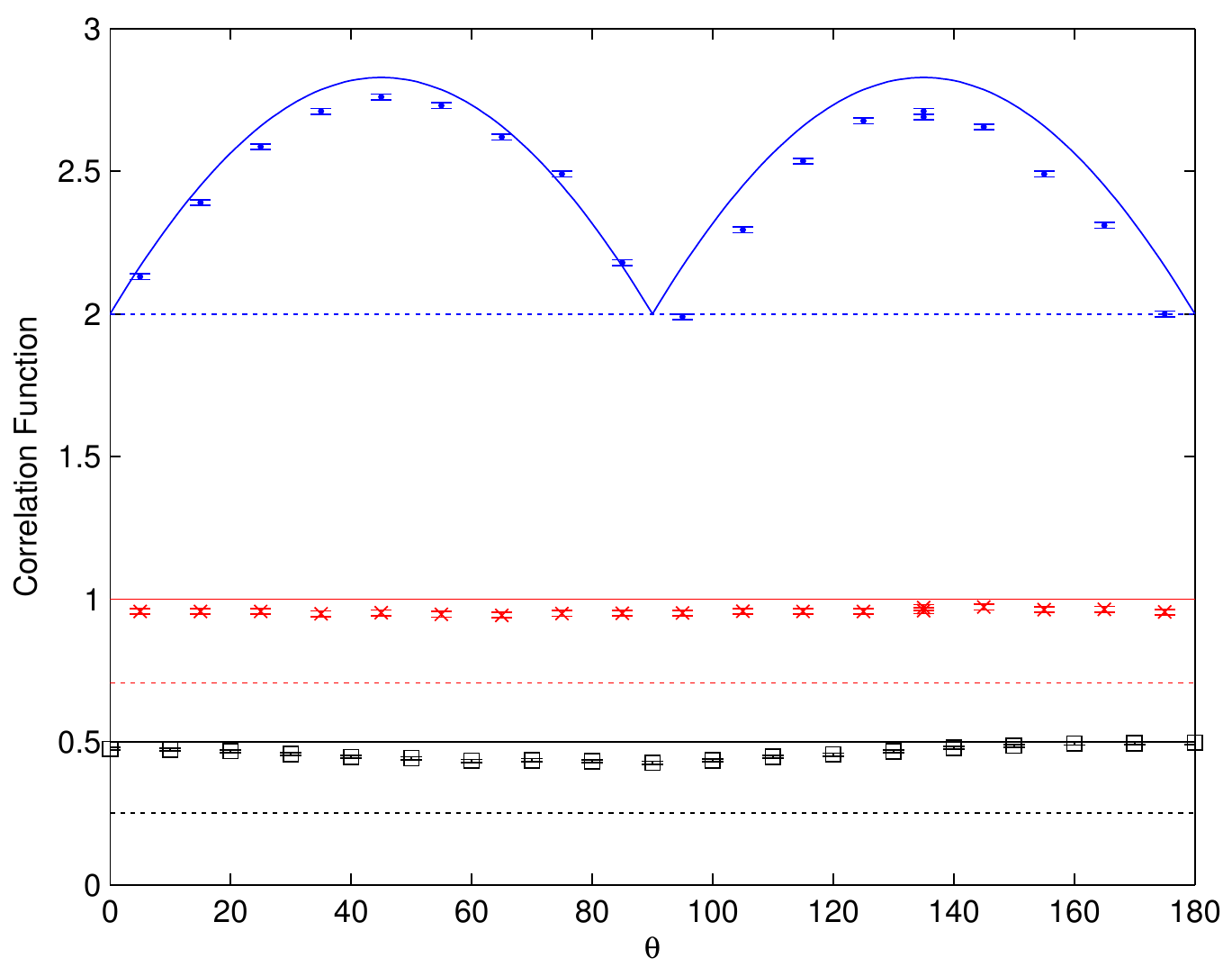}\end{center}
\vspace{-4ex} \caption{Alignment dependence of nonlocality measures (Color online). The top (blue), middle (red) and bottom (black) data and curves show the dependence on $\theta$ (Fig. 1) of the CHSH, EPR-steering and entanglement tests respectively. Solid lines are the theoretical values for a perfect singlet state, the dashed lines represent the bounds for each non-local task, and the points are experimentally determined values.}\label{fig:Data}\end{figure}

The results for each correlation function, as $\theta$ varies, can be seen in Fig.\ \ref{fig:Data}. The inequality for the CHSH test was varied to take into account the effective relabelling of the measurement outcomes every ${\pi}/{2}$. To do this, we replace the right-hand-side of \erf{CHSH} with $\rm{max}\left(|\rm{perm}([-1,1,1,1])\cdot[AB,A'B,AB',A'B']|\right)$. This ensures that the CHSH test is violated for almost all values of $\theta$, but the degree of violation depends strongly upon $\theta$. By contrast, the 
EPR-steering and separability demonstrations are observed to be  almost completely rotationally invariant.  
The systematic deviations from the predicted complete rotational invariance arise predominantly from imperfections in symmetrically implementing the rotated trine measurement. Slightly imperfect splitting ratios (errors of about 1\% to 2\%) of the partially polarizing beam splitters makes the experimental trine measurement slightly asymmetric. Additionally, the slightly imperfect retardance of the half-wave plate ($\theta^\prime$ in Fig.~2) tilts the trine measurement out of the desired plane in Bloch space in a setting-dependent manner. 

\section{Conclusion} 
 
We have shown that the complexity cost, $W$, of the nonlocal phenomena of Bell nonlocality, EPR-steering and entanglement form a strict hierarchy ($W_{\rm B}>W_{\rm S}>W_{\rm E}$), reflecting the strength of the concept of nonlocality being tested in each case \cite{Wiseman2007}. We have done this by introducing new inequalities for EPR-steering and entanglement 
that are the simplest possible \textit{witnesses} for these types of nonlocality. That is, a positive result confirms the effect has been seen. Note however that a negative result does not necessarily mean the effect is not present  in our experiment, as a different inequality (using the same detector settings)  might detect it.
 Moreover, for particular classes of states it is likely that one could design equally parsimonious tests, using different measurement settings, which would allow demonstrations of EPR-steering and entanglement when our measurement settings do not. None of this alters the key point that no states or measurement settings allow for tests that are more parsimonious than those we have derived and demonstrated. 

 The tests we introduce for EPR-steering and entanglement could  have practical application in entanglement distribution. By using 
 the minimum number of joint measurements settings, and the minimum number of joint outcomes, our tests could maximize the rate at which 
 these tasks are performed. The fact that the degree of violation is independent of the relative orientation of the two polarizers could have technological advantages in implementations using optical fibers. 
It is also of theoretical interest, showing the application of the concepts of 2-designs to quantum nonlocality.  \\

{\bf Acknowledgment}: Part of  this research was conducted by the Australian Research Council Centre of Excellence for Quantum Computation and Communication Technology (project number CE110001029). SMB thanks the Royal Society and the Wolfson Foundation for support.

\appendix \section{Measurement efficiency}

The different measurement outcomes have slightly different efficiencies due to small experimental imperfections. These measurement channel imperfections, at the level of a few percent or less, include unit-to-unit variation in the efficiencies of the single photon counting modules, and slight imbalances in fiber coupling efficiencies. The different efficiencies of the measurement outcomes could lead to errors in the correlation function if not compensated. We fix this problem by carefully measuring the relative efficiencies (i.e. the ratios of measurement channel efficiencies). We post-process the measurement data to reduce the efficiency in high-efficiency channels so that all relative measurement efficiencies are effectively the same. 

The ratio of channel efficiency is defined as $\eta_{nm}=\eta_n/\eta_m$, where $\eta$ is the efficiency after separation on a (P)PBS into output channels $n$ and $m$. The correction factor $\eta_{nm}$ can be calculated by measuring both $\s{x}$ and $-\s{x}$, by rotating a waveplate before the splitting device, and comparing the count rates after the eigenvectors switch channels. To apply the correction, we take the raw counts $C_n$ and $C_m$ for a particular setting, then multiply $C_m$ by $\eta_{nm}$ to correct for any asymmetric loss. This effectively gives both channels the same loss, $\eta_n$.  This calibration technique allows for a demonstration of the non-locality tests whilst minimizing the systematic error from unbalanced channel efficiencies. The observed channel efficiency ratios varied between 0.9 and 1.1 for all combinations of the 6 different output channels.

\end{document}